# Scale-free statistics of time interval between successive earthquakes

Sumiyoshi Abe[a,]*, Norikazu Suzuki[b]

[a]*Institute of Physics, University of Tsukuba, Ibaraki 305-8571, Japan*
[b]*College of Science and Technology, Nihon University, Chiba 274-8501, Japan*

**Abstract**   The statistical property of the calm times, i.e., time intervals between successive earthquakes with arbitrary values of magnitude, is studied by analyzing the seismic time series data in California and Japan. It is found that the calm times obey the Zipf-Mandelbrot power law, exhibiting a new scale-free nature of the earthquake phenomenon. Dependence of the exponent of the power-law distribution on threshold for magnitude is examined. As threshold increases, the tail of the distribution tends to become longer, showing difficulty in statistically estimating time intervals of earthquakes.

___

*Corresponding author. FAX: 81-29-853-4312.

*E-mail address:* suabe@sf6.so-net.ne.jp (S. Abe)

# 1. Introduction

Although seismicity is characterized by complex phenomenology that makes it difficult to develop coherent explanation and prediction of earthquakes, some known empirical laws are remarkably simple. Classical examples are the Omori law [1] for temporal pattern of aftershocks and the Gutenberg-Richter law [2] for frequency and magnitude. These are power laws and represent the scale-free natures of the earthquake phenomenon. There is also an attempt to unify the Omori law and the Gutenberg-Richter law within a single hypothetical scaling law [3].

One of the extreme goals of seismology is to predict when and where the next main shock will occur after an earlier main shock. Though science seems still far from such a goal, there are some attempts based on peculiar properties of precursory phenomena (Refs. [4,5] and the references threrein). Recently, we have analyzed the seismic data from the viewpoint of science of complexity. First, we have studied the statistical properties of the three-dimensional distance between successive earthquakes [6] and have discovered that the distance follows the "$q$-exponential distribution" with $0 < q < 1$ in nonextensive statistical mechanics [7], which maximizes the Tsallis entropy [8] under appropriate constraints (see Sec. 2). Then, we have constructed a network associated with earthquakes and have shown [9-11] that it is an evolving scale-free network (see also Ref. [12]). Furthermore, we have investigated the nonstationary feature of the Omori regime, in which the Omori law holds, and have found [13] that the aging phenomenon is exhibited and the scaling law can be established for aftershocks, suggesting a possible relevance of glassy dynamics to the mechanism of aftershocks.

In this paper, we report the discovery of a new scale-free nature in the earthquake phenomenon. We analyze the seismic data taken in California and Japan, and study the statistical property of the time interval between two successive earthquake events with arbitrary values of magnitude. Such an interval is referred to here as the "calm time". A collection of the calm times thus obtained gives a histogram, from which the cumulative waiting time distribution can be calculated. We show that statistics of the calm time is nicely characterized by the distribution of the Zipf-Mandelbrot type [14], which corresponds to the $q$-exponential distribution with $q > 1$, exhibiting a novel scale-free nature of the earthquake phenomenon. The present result combined with the one in Ref. [6] may feature spatio-temporal complexity of seismicity in a novel manner. We shall also discuss similarities between earthquakes and the Internet in their temporal behaviors in view of a common complex-network structure.

The paper is organized as follows. In Sec. 2, we present a brief discussion about the Zipf-Mandelbrot law within the framework of nonextensive statistical mechanics, to make the presentation self-contained. In Sec. 3 and Sec. 4, we report our results on statistical properties of the calm times in California and Japan, respectively. In Sec. 5, we examine the influence of threshold for magnitude on statistics by employing the data in California. Sec. 6 is devoted to concluding remarks.

## 2. Zipf-Mandelbrot law

As will be seen in the subsequent sections, the calm time, $\tau$, obeys the cumulative distribution of the Zipf-Mandelbrot type

$$P(>\tau) = \frac{1}{(1+\varepsilon\tau)^{\alpha}} \qquad (\alpha, \varepsilon > 0). \tag{1}$$

We rewrite Eq. (1) in the following form:

$$P(>\tau) = e_q(-\tau/\tau_0), \tag{2}$$

where $e_q(x)$ stands for the $q$-exponential function defined by

$$e_q(x) = [1+(1-q)x]_+^{1/(1-q)} \qquad ([a]_+ \equiv \max\{0, a\}). \tag{3}$$

For the later purpose, here we also mention the inverse function of the $q$-exponential function, which is termed the "$q$-logarithmic function" and is given by

$$\ln_q(x) = \frac{1}{1-q}(x^{1-q} - 1). \tag{4}$$

In the limit $q \to 1$, $e_q(x)$ and $\ln_q(x)$ tend to the ordinary exponential and logarithmic functions, respectively.

In the representations in Eqs. (1) and (2), $\alpha$, $\varepsilon$, $q$ and $\tau_0$ are related to each other as $\alpha = 1/(q-1)$ and $\varepsilon = (q-1)/\tau_0$. The Zipf-Mandelbrot distribution decays as a power law, corresponding to the case

$$q > 1. \tag{5}$$

It is known [7,15] that the $q$-exponential distribution can be derived from the maximum entropy principle with the Tsallis entropy [8]

$$S_q[p] = \frac{1}{1-q}\left(\int \frac{d\tau}{\sigma}[\sigma p(\tau)]^q - 1\right) \tag{6}$$

where $q$ and $\sigma$ are the positive entropic index and a scale factor with the dimension of time, respectively. In the limit $q \to 1$, it converges to the Boltzmann-Gibbs-Shannon entropy $S[p] = -\int d\tau\, p(\tau) \ln[\sigma p(\tau)]$. Under the constraints on normalization of the "original distribution"

$$\int d\tau\, p(\tau) = 1 \tag{7}$$

and on the generalized expectation value

$$<\tau>_q = \int d\tau\, \tau\, P_q(\tau) \tag{8}$$

with the "escort distribution" [16,17] given by

$$P_q(\tau) = \frac{[p(\tau)]^q}{\int d\tau'\, [p(\tau')]^q}, \tag{9}$$

the Tsallis entropy in Eq. (6) is maximized by the following stationary distribution:

$$\tilde{p}(\tau) = \frac{1}{Z_q} e_q(-\lambda^*(\tau - <\tau>_q)), \qquad (10)$$

where

$$Z_q = \int_0^{\tau_{max}} d\tau\, e_q(-\lambda^*(\tau - <\tau>_q)). \qquad (11)$$

Here, $\lambda^*$ is related to the Lagrange multiplier, $\lambda$, associated with the constraint on the generalized expectation value in Eq. (8), as follows:

$$\lambda^* = \frac{\lambda}{\int_0^{\tau_{max}} \frac{d\tau}{\sigma} [\sigma \tilde{p}(\tau)]^q}. \qquad (12)$$

The upper bound in the above integrals are: $\tau_{max} \to \infty$ if $q \geq 1$, whereas $\tau_{max} = <\tau>_q + [(1-q)\lambda^*]^{-1}$ if $0 < q < 1$. (As already noted, the Zipf-Mandelbrot distribution corresponds to the former case $q > 1$.) An important point in this formalism is that the quantity to be compared with the observed distribution is not the original distribution but the escort distribution [18]. Therefore, the cumulative distribution should be defined as follows:

$$P(>\tau) = \int_\tau^{\tau_{max}} d\tau'\, \tilde{P}_q(\tau'), \qquad (13)$$

where $\tilde{P}_q$ is the escort distribution associated with the original stationary distribution, $\tilde{p}$, in Eq. (10). After some calculations, this cumulative distribution is found to be given by the one in Eq. (2) with

$$\tau_0 = (1-q)<\tau>_q + \frac{1}{\lambda^*}, \qquad (14)$$

which is always positive [19].

We point out that there are some known approaches to derive the Zipf-Mandelbrot distribution other than the above-mentioned maximum Tsallis-entropy principle. However, Tsallis statistics in fact plays a distinguished role in describing some aspects of seismicity. The distribution of the spatial distance between successive earthquakes is found to obey the Tsallis $q$-exponential distribution with $q<1$ [6], which corresponds to the Zipf-Mandelbrot distribution with a *negative exponent* and does not seem to be described in simple manners other than the maximum Tsallis-entropy principle. It is our hope to understand aspects of seismicity as an example of complex phenomena in a unified manner, and therefore we take special notice of a role of the Tsallis entropy.

3. **Calm-time distribution in California**

Here, we present the result of data analysis of the earthquake catalog made available by the Southern California Earthquake Data Center (http://www.scecdc.scec.org/

catalogs.html) covering the period between 00:25:8.58 on January 1, 1984 and 15:23:54.73 on December 31, 2002 in the region spanning $29°15.25' N - 38°49.02' N$ latitude and $113°09.00' W - 122°23.55' W$ longitude. (We have taken this period since the data in 1983 are partially missing for a few months.) The total number of events is 375887. The used data downloaded on February 13, 2003 contained not only strong earthquakes but also extremely weak earthquakes like magnitude 0.0.

Fig. 1 (a) and (b) show the log-log and semi-$q$-log plots of the cumulative distribution of the calm time, respectively. The dots represent the observed data, whereas the solid line in Fig. 1 (b) corresponds to the model in Eq. (2). One can appreciate that the observed data are nicely described by the Zipf-Mandelbrot law in the $q$-exponential form.

## 4. Calm-time distribution in Japan

It is of interest to examine the result for the data in California by comparing other ones. In this section, we present the analysis of the catalog of earthquakes taken in Japan, which is made available by the Japan University Network Earthquake Catalog (http://kea.eri.u-tokyo.ac.jp/CATALOG/junec/monthly.html) covering the period between 01:14:57.63 on January 1, 1993 and 20:54:38.95 on December 31, 1998 in the region spanning $25.851° N - 47.831° N$ latitude and $126.433° E - 148.000° E$ longitude. The number of events is 123390. (We have limited ourselves to this period since before 1993 the number of the observed data per year turned out to be about half of the latter period, indicating incompleteness of the database. An essential difference of the catalog of the Japan University Network Earthquake Catalog from that of the Southern California Earthquake Data Center is that the former ones contain threshold for the value of magnitude, 2, unfortunately.)

Similarly to the previous section, Fig. 2 (a) and (b) show the log-log and semi-$q$-log plots of the cumulative distribution of the calm time, respectively. The dots represent the observed data, whereas the solid line in Fig. 1 (b) corresponds to the model in Eq. (2). As in the case of California, again one may appreciate that the observed data are well described by the Zipf-Mandelbrot law in the $q$-exponential form.

**5. Dependence on threshold for magnitude**

As mentioned in Section 4, the Japanese data contain threshold for the value of magnitude, but the data are still described well by the Zipf-Mandelbrot law. This fact immediately leads to the following question: How the feature of the distribution depends on threshold for magnitude? In other words, what is the time interval between successive earthquakes larger than a given value of magnitude? We have performed such analysis by making use of the data in California employed in Section 3. The result we found is that, up to a certain value of magnitude, the calm-time distribution remains as the Zipf-Mandelbrot type.

In Fig. 3, we present the pattern of variations of $q$ and $\tau_0$ on threshold for magnitude, $m_{th}$. There, we find that the value of $q$ monotonically increases with respect to the value of threshold. This means that the tail of the distribution tends to become heavier for a higher value of threshold, implying difficulty in statistically estimating time intervals of earthquakes.

**6. Concluding remarks**

We have discovered a new scale-free nature of the earthquake phenomenon, in connection with the concept of the calm time between successive earthquakes. We have observed that the calm times in California and Japan well obey the Zipf-Mandelbrot law represented by the $q$-exponential distribution with $q > 1$. We have also examined how the two parameters contained in the $q$-exponential distribution, $q$ and $\tau_0$, depend on threshold for magnitude. These results combined with the spatial properties reported in Ref. [6] feature spatio-temporal complexity of seismicity in a novel manner. An important point arising here is that this fact enables us to develop a thermodynamics-like approach to seismicity and should provide new insights into the study of earthquakes. This is because Tsallis statistics is known to be consistent with the principles of thermodynamics.

In the present study, we have considered rather long time trends of the seismic time series. However, it turned out [20] that there exists the *nested structure* in statistics of the calm time. On the relatively shorter time scale, one observes transition between the $q$-exponential distributions, i.e., $(q_1, \tau_{0,1}) \rightarrow (q_2, \tau_{0,2}) \rightarrow (q_3, \tau_{0,3}) \rightarrow \cdots$. The points of transition appear to be corresponding to mainshocks. In an interval between transitions, statistics of the calm time is characterized by the $q$-exponential distribution, which is a stationary distribution maximizing the Tsallis entropy. Between such stationary states, nonstationary intervals are identified. According to our analysis [20], such nonstationary regimes are the time intervals after mainshocks followed by swarms of aftershocks, in which the Omori law [1] for aftershocks holds. Remarkably, this feature is highly analogous to that found in our recent study of the Internet traffic. There, we have performed the Ping experiments in order to examine the state of congestion of the Internet and have collected the time series data of the round-trip time of the Ping signal. The sparseness time, which is the time interval between the successive round-trip time above a fixed value of threshold, corresponds to the calm time the seismic time series. In Ref. [21], we have found that the Internet itinerates over a series of stationary states, which are all described by the maximum-Tsallis-entropy distributions, i.e., the $q$-exponential distributions. In addition, we have found that both the Omori law [22] and the Gutenberg-Richter law [23] also hold for the Internet time series, where seismic magnitude corresponds to (the logarithm of) the round-trip time of the Ping signal. These striking similarities between seismicity and the Internet may have an origin in the structure of complex networks with scale-free topology [9-12,24,25].

**Acknowledgment**

The authors thank the support by the Grant-in-Aid for Scientific Research of Japan Society for the Promotion of Science.

# Figure Captions

**Fig. 1** (a) The log-log plot of the cumulative distribution of the calm times in California. The histogram is made by using the equal time interval of 10 [$s$]. No threshold is set. (b) The corresponding semi-$q$-log plot of the cumulative distribution of the calm time in California. The solid line represents the model in Eq. (2). The best-fit regression is realized by $q = 1.13$ and $\tau_0 = 1.724 \times 10^3$ [$s$]. The associated value of the correlation coefficient between the data and the model (i.e., the straight line) is $\rho = -0.98828$.

**Fig. 2** (a) The log-log plot of the cumulative distributions of the calm times in Japan. The histogram is made by using the equal time interval of 10 [$s$]. The value of threshold for magnitude is 2. (b) The corresponding semi-$q$-log plot of the cumulative distribution of the calm time. The solid line represents the model in Eq. (2). The best-fit regression is realized by $q = 1.05$ and $\tau_0 = 1.587 \times 10^3$ [$s$]. The associated value of the correlation coefficient between the data and the model (i.e., the straight line) is $\rho = -0.99007$.

**Fig. 3** Dependency of $(q, \tau_0)$ on the value of threshold for magnitude. The data taken in California is employed. From the bottom to the top, $m_{th}$ = 0.0, 1.4, 2.0, 2.1, 2.2, 2.3, 2.4, 2.5. The values of the correlation coefficient are larger than $\rho = -0.96341$.

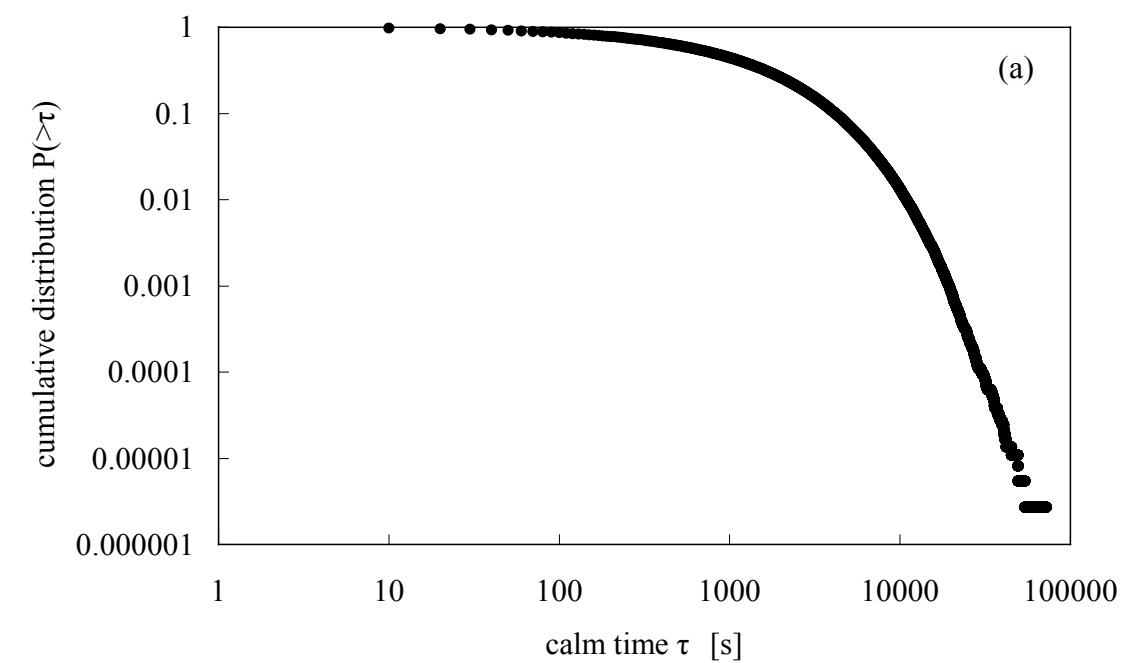

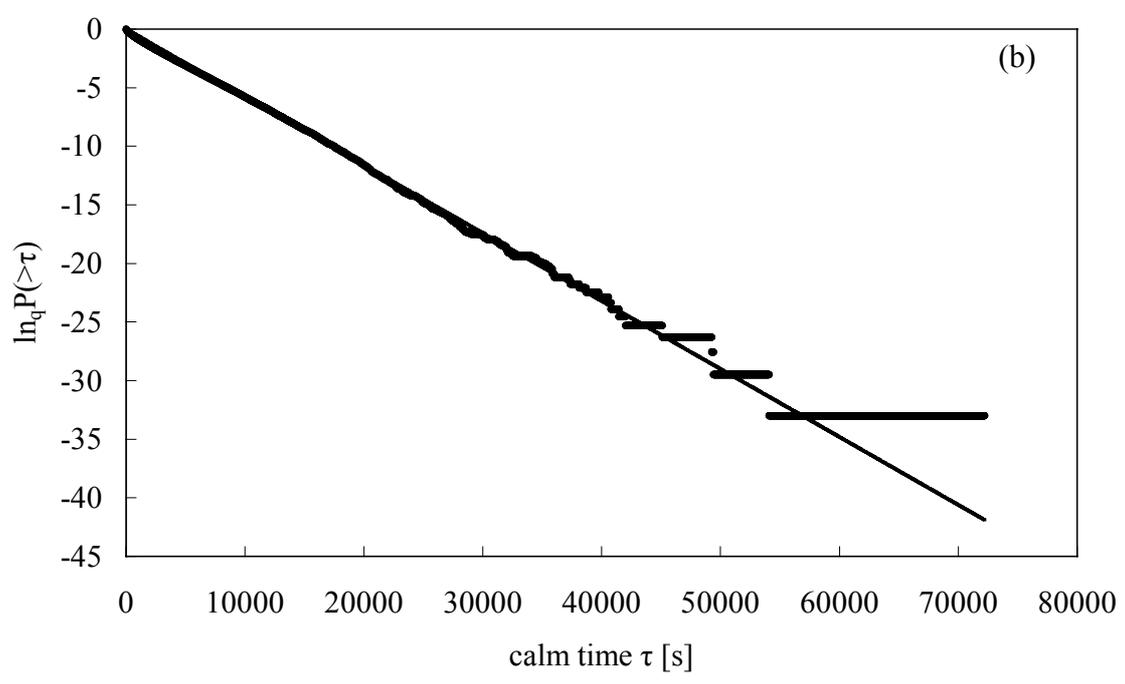

Fig. 1

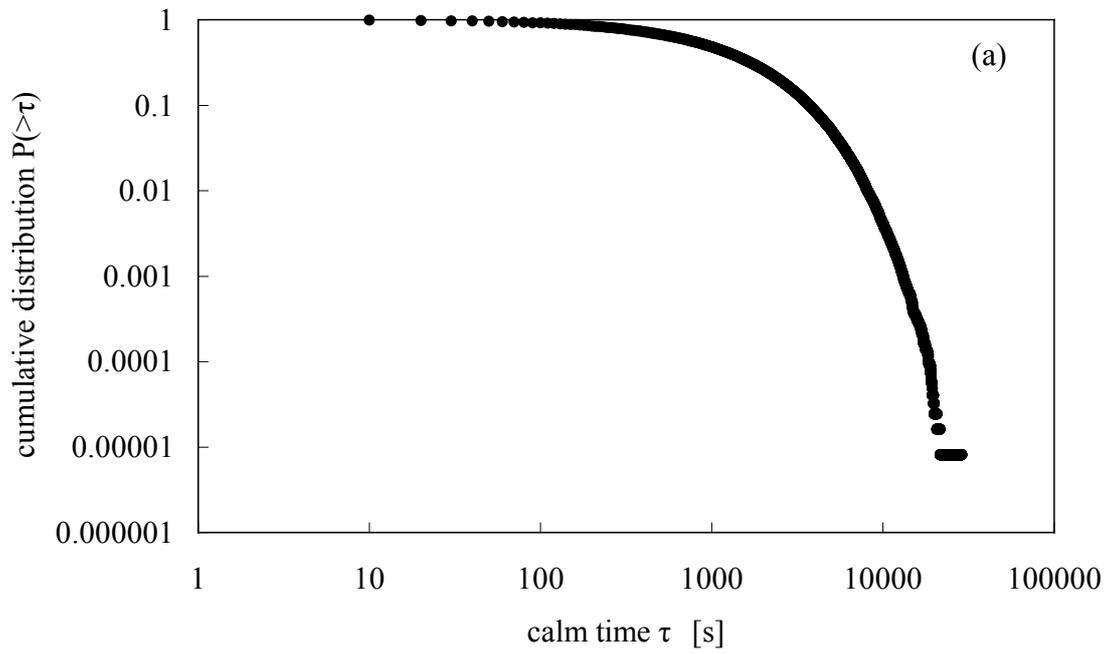

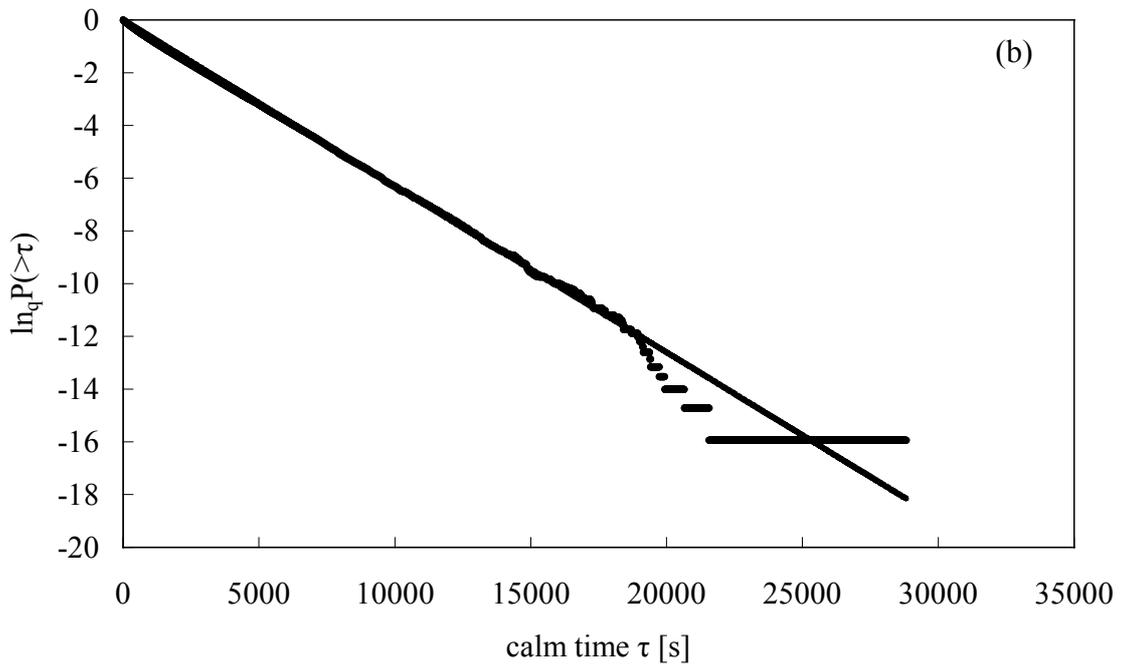

Fig. 2

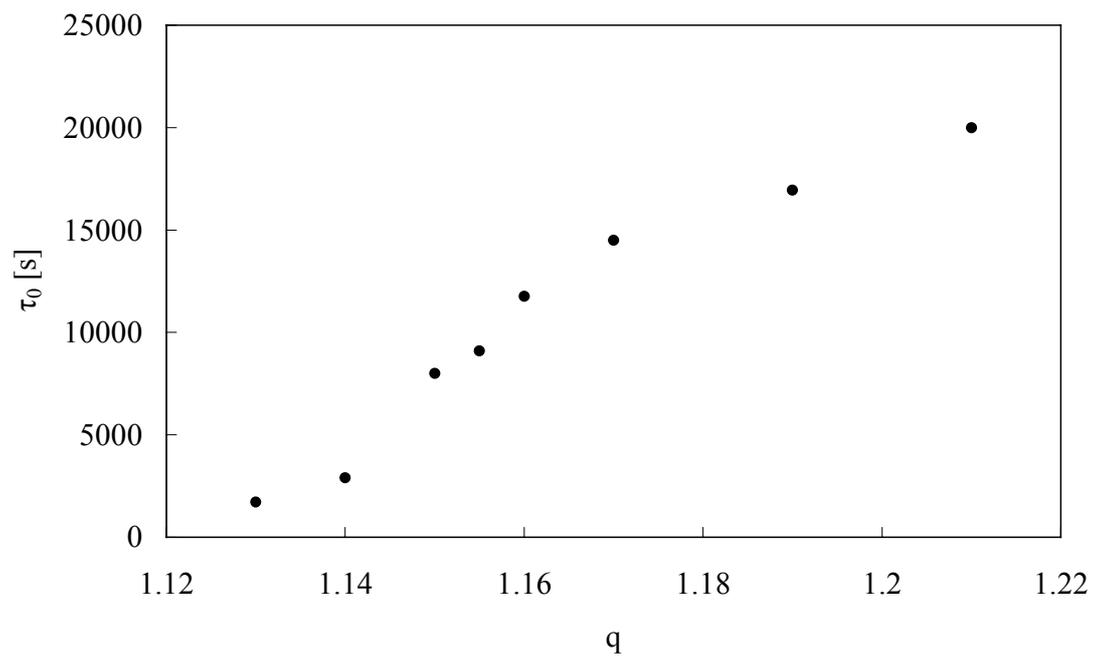

Fig. 3